\documentclass[prb,aps,twocolumn,showkeys,showpacs,floatfix,superscriptaddress]{revtex4-1}
\usepackage{graphicx}
\usepackage{hyperref}
\usepackage{amsmath}
\usepackage{amssymb}
\usepackage{natbib}

\begin{document}
\title{Low-amplitude magnetic vortex core reversal by non-linear interference between azimuthal spin waves and the vortex gyromode.}
\author{Markus Sproll}
\author{Matthias Noske}
\affiliation{Max Planck Institute for Intelligent Systems (formerly MPI for Metals Research)\\ Heisenbergstr. 3, 70569 Stuttgart, Germany}
\author{Hans Bauer}
\affiliation{University of Regensburg, Department of Physics \\Universit\"atsstr. 31, 93040 Regensburg, Germany}
\author{Matthias Kammerer}
\affiliation{Max Planck Institute for Intelligent Systems (formerly MPI for Metals Research)\\ Heisenbergstr. 3, 70569 Stuttgart, Germany}
\author{Ajay Gangwar}
\affiliation{University of Regensburg, Department of Physics \\Universit\"atsstr. 31, 93040 Regensburg, Germany}
\author{Georg Dieterle}
\author{Markus Weigand}
\author{Hermann Stoll}
\affiliation{Max Planck Institute for Intelligent Systems (formerly MPI for Metals Research)\\ Heisenbergstr. 3, 70569 Stuttgart, Germany}
\author{Georg Woltersdorf}
\author{Christian H. Back} 
\affiliation{University of Regensburg, Department of Physics \\Universit\"atsstr. 31, 93040 Regensburg, Germany}
\author{Gisela Sch\"utz}
\affiliation{Max Planck Institute for Intelligent Systems (formerly MPI for Metals Research)\\ Heisenbergstr. 3, 70569 Stuttgart, Germany}

\begin{abstract}

We demonstrate a non-linear interference due to an active 'dual frequency' excitation of both, the sub-GHz vortex gyromode and multi-GHz magneto-static spin waves in ferromagnetic micro\-meter sized platelets in the vortex state. When the sub-GHz vortex gyromode is excited simultaneously a significant broadband reduction of the switching threshold for spin wave mediated vortex core reversal is observed in both, experiments and micromagnetic simulations. Consequently, the magnetic field amplitudes required for vortex core reversal can be lowered by nearly one order of magnitude. Moreover, additional spin wave resonance frequencies are found which emerge only if the vortex gyromode is actively excited simultaneously which can be explained by frequency doubling and by the broken symmetry of the vortex state.
\end{abstract}
\pacs{75.78.-n, 75.25.-j, 75.30.Ds, 75.75.-c}
\keywords{magnetic vortex dynamics, vortex core reversal, spin wave, vortex gyromode, frequency doubling, non-linear interference}
\maketitle

\section{INTRODUCTION}
The magnetic vortex is the simplest, non trivial ground state configuration of micron-sized soft magnetic thin film platelets with an in-plane curling magnetization and a perpendicularly magnetized core pointing either up (polarization $p$ = +1) or down ($p$ = -1) \cite{ShinjoScience,WachowiakScience,ArgylePRL,GuslienkoJApplPhys}. Essential progress in the understanding of nonlinear vortex dynamics was achieved when low-field core toggling was discovered by excitation of the gyrotropic eigenmode at sub-GHz frequencies by linear in-plane ac magnetic fields \cite{BartelNature}. The switching mechanism is explained by the creation and subsequent annihilation of a vortex-antivortex (VA) pair \cite{BartelNature,HertelPRL,XiaoAPL,GuslienkoPRL,ArneNature}. This switching mechanism has been proven to be almost universal and independent of the type of excitation, e.g., by alternating \cite{BartelNature}, rotating \cite{CurcicPRL} or pulsed \cite{WeigandPRL} magnetic fields or by spin polarized currents \cite{YamadaNatMater,YamadaAPL}. 

At frequencies more than an order of magnitude higher compared to the gyrotropic eigenfrequency, vortex structures possess spin wave eigenmodes arising from the magneto-static interaction, characterized by their radial ($n$=1, 2, 3,...) and azimuthal ($m$=0, $\pm$1, $\pm$2, $\pm$3,...) mode numbers. The interaction between the vortex core and the azimuthal spin waves lifts the degeneracy of the rotating clockwise (CW, corresponding to $m<1$) and anti-clockwise (CCW, $m>1$) modes leading to an observable frequency splitting \cite{IvanovPRB,ZaspelPRB,ParkPRL,ZhuPRB,HoffmannPRB,GuslienkoPRL2,GuslienkoPRB}. Recently we demonstrated experimentally and by micromagnetic simulations that a significant faster vortex core reversal can be achieved by exciting the azimuthal ($n$=1, 2, ..., $m$=$\pm$1) spin wave modes with in-plane rotating magnetic fields in the multi-GHz frequency range \cite{KammererNature,KammererPRB,KammererAPL}. Consequently, in the same way as demonstrated for sub-GHz gyromode excitation \cite{CurcicPRL,CurcicPSSb}, unequivocal and unidirectional spin wave mediated vortex core switching to its 'up' or 'down' polarization can be achieved by tuning the frequency and sense of rotation of a multi-GHz in-plane rotating magnetic field to a spin-wave eigenmode \cite{KammererNature}. When switching the vortex core in platelets of 1.6~$\mu$m in diameter with bursts of 24 periods, applied fields of less than 1~mT are necessary, resulting in a relatively long excitation time of approximately 6~ns. In contrast, the application of a rotating 'one-period' field burst results in much faster switching times down to 220 ps \cite{KammererPRB} but at the cost of a significant larger threshold amplitude of 5.0~mT to 5.5~mT (cf. Fig.~4 in \cite{KammererPRB}). From technological aspects, e.g., for a possible realization of a V(ortex)C(ore)-MRAM even shorter switching times are desired. It was already demonstrated that a decrease of the vortex core switching time is limited by the time needed for the energy transport of the homogeneous distributed excitation energy to the centre of the sample \cite{KammererAPL}. Thus smaller platelets are necessary for obtaining shorter vortex core switching times. However, the threshold amplitude for vortex core reversal is enhanced significantly in smaller samples to values between 30~mT and 100~mT \cite{KammerertPhD}, reaching the limit of magnetic field strengths which can be created by striplines. Therefore, it is of significant technological interest to find a possibility for lowering the switching threshold amplitudes even at shorter switching times.

In the present work we studied an active 'dual frequency' excitation of both (i) the sub-GHz vortex gyromode and (ii) multi-GHz magneto-static spin waves as shown in Fig.~\ref{Fig1}. In the observed frequency range (2.0~GHz to 5.5~GHz, c.f. Fig.~\ref{Fig2}), non-linear interference effects are causing a broadband reduction of the energy input of up to almost one order of magnitude for the spin-wave mediated switching threshold. Furthermore we obtained the appearance of additional spin wave eigenmodes. Micromagnetic simulations with a static displacement of the vortex core show that the symmetry break due to a shifted vortex core is not responsible for these effects as they are only present when the vortex core is shifted dynamically from its equilibrium position in the center of the disc by an active excitation of the gyromode. We present experimental results, micromagnetic simulations and give an explanation of the observed effects emerging from the interplay of azimuthal spin waves with the vortex gyromode.

\section{METHODS}
\begin{figure}
    \centerline{\includegraphics[clip,width=0.45\textwidth]{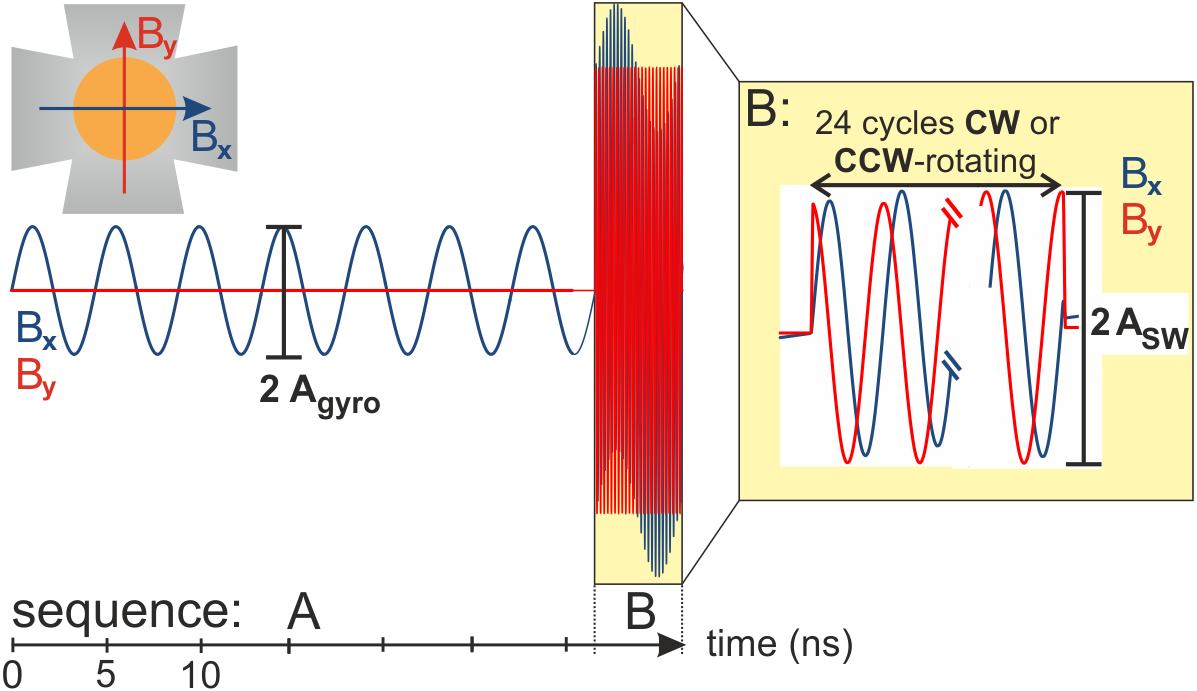}}
    \caption{Scheme of the simultaneous gyromode and spin wave excitation. A linear low frequency excitation field with varying amplitude is applied along the x direction for 7 periods followed by a 24 period multi-GHz rotating field burst with variable amplitude and frequency. 
    }
    \label{Fig1}
\end{figure}

Experiments were performed on ferromagnetic Permalloy (Py) discs in the vortex magnetic ground state (1.6~$\mu$m diameter, 50~nm thick) prepared by e-beam lithographie and metal deposition on a silicon nitride membrane. As shown in Fig.~\ref{Fig1} we used two crossed copper striplines below the Py disc to excite in-plane rotating magnetic fields \cite{CurcicPRL,CurcicPSSb}. The out-of-plane magnetization of the vortex core was recorded at the Ni L$_3$ edge before and after each excitation by magnetic scanning transmission soft X-ray microscopy at the MAXYMUS endstation (BESSY II, Helmholtz Zentrum Berlin) which provides a spatial resolution of 25~nm (determined by the Fresnel zone plate) and for time-resolved stroboscopic pump-and-probe measurements a time resolution of 40~ps, given by the time spread of the X-ray flashes. In this way vortex core polarity switching was detected experimentally.

\section{EXPERIMENTAL RESULTS}
The combined excitation scheme used in our experiments is sketched in Fig.~\ref{Fig1}: A linear (sub-threshold) gyromode excitation was applied along the x direction (sequence A in Fig.~\ref{Fig1}) to achieve a steady state motion of the vortex core with a nearly constant gyration radius which is approximately reached after 7 periods. Subsequently an additional in-plane rotating multi-GHz field with 24 periods duration was used for spin wave excitation in order to induce the vortex core reversal (sequence B in Fig.~\ref{Fig1}). The spin wave mediated switching threshold was measured for different amplitudes $A_{\text{gyro}}$ but with fixed frequency $f_{\text{gyro}}=185$~MHz (corresponding to the measured resonance frequency of the gyromode) for the linear sub-GHz gyromode excitation (Fig.~\ref{Fig1}: sequence A) by varying both frequency $f_{\text{SW}}$ and amplitude $A_{\text{SW}}$ of the multi-GHz in-plane rotating field burst (Fig.~\ref{Fig1}: sequence B). The phase relation between gyromode- and GHz-excitation was always kept constant.
\begin{figure}
    \centerline{\includegraphics[clip,width=0.45\textwidth]{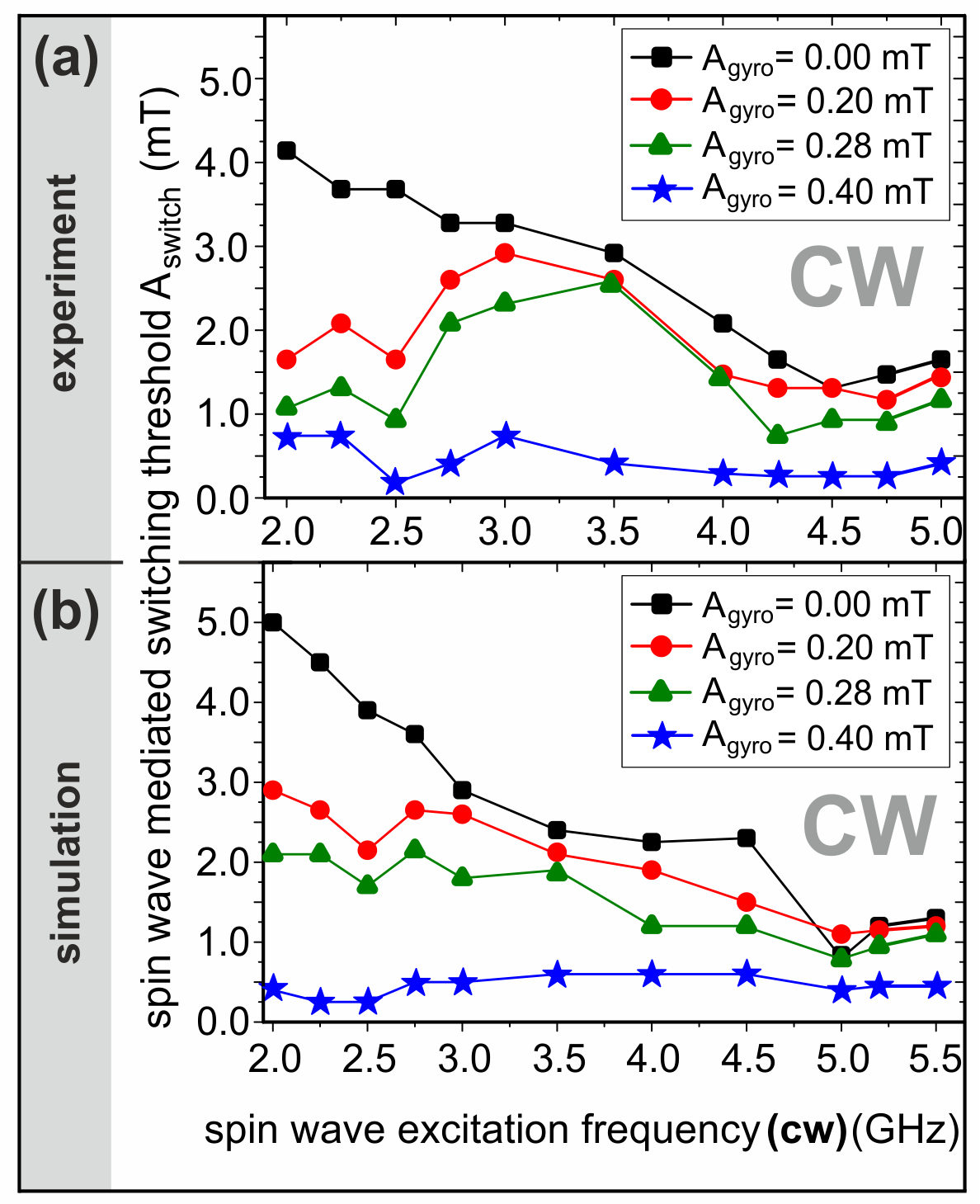}}
    \caption{Experimental results (a) and micromagnetic simulations (b) of the vortex core switching threshold vs. spin wave excitation frequency (CW sense of rotation). Distinct switching thresholds are shown for varying amplitudes of a simultaneous linear (sub-GHz) gyromode excitation ($f_{\text{gyro,exp}}=185$~MHz, $f_{\text{gyro,sim}}=240$~MHz). For more details see text.
    }
    \label{Fig2}
\end{figure}
In Fig.~\ref{Fig2}(a) (experiments) and \ref{Fig2}(b) (micromagnetic simulations) the switching thresholds for CW rotating spin waves are displayed as a function of frequency. The initial state in all experiments was a vortex core up ($p$ = 1) and all applied linear gyromode amplitudes $A_{\text{gyro}}$ were chosen below the switching threshold (0.48~mT) in order to ensure that the reversal was induced only by the additional spin wave excitation. The black squares represent the switching threshold without additional gyromode excitation, the other symbols show the spin wave induced switching threshold for different simultaneously applied linear gyromode amplitudes. Without additional gyromode excitation (black squares) both experimental data and micromagnetic simulations reveal a minimum at the expected 4.8~GHz which corresponds to the ($n$=1, $m$=-1) spin wave eigenmode as described in detail in \cite{KammererNature}. As in \cite{KammererNature}, no additional resonances in the switching threshold were found below 4.8~GHz for $A_{\text{gyro}}=0.00$~mT. Note that the sample geometry used in the present paper is in fact the same as in \cite{KammererNature}. The first  phenomenon arising from the actively applied combined gyromode spin wave excitation observed in Fig.~\ref{Fig2}(a) is a broadband reduction of the switching threshold with increasing gyromode amplitude. This decrease reaches large values of up to approximately one order of magnitude at 2~GHz with the highest applied gyromode amplitude. Even at the spin wave resonance frequency where the lowest switching threshold without additional gyromode excitation can be found, the amplitude is reduced by a factor of 6. In addition to this broadband reduction, the switching thresholds show an additional local minima at 2.5~GHz for simultaneously applied gyromode excitation.

\section{MICROMAGNETIC SIMULATIONS}
To confirm these experimental results, micromagnetic simulations using the OOMMF code \cite{OOMMF} were performed, shown in Fig.~\ref{Fig2}(b), with sample dimensions and initial states as in the experiments (diameter: 1.6~$\mu$m, thickness: 50~nm, initial vortex core: up, switching induced by CW rotating spin waves, other simulation parameters as in \cite{KammererNature}). The two effects found experimentally, the broadband decrease and the minimum of the switching threshold at 2.5~GHz are qualitatively well reproduced by the simulations. The differences in the spin wave switching amplitudes and resonance frequencies between simulations and experiments as observed also in preceding studies \cite{KammererNature,KammererPRB,ArneNJPhys} can be explained by slight differences between the experimental and the simulated sample parameters as well as by defects resulting from sample preparation and the influence of the surface roughness of the measured sample \cite{ArneNJPhys} compared to the perfect surface in our simulations.
\begin{figure}
    \centerline{\includegraphics[clip,width=0.45\textwidth]{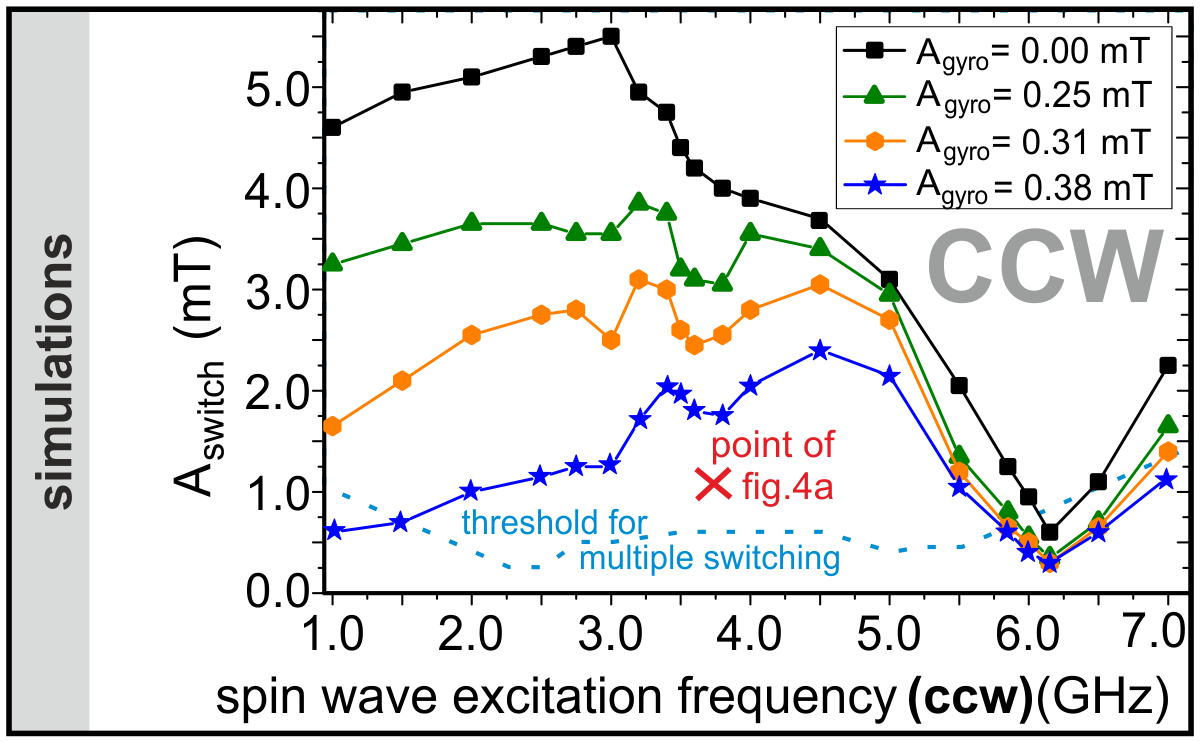}}
    \caption{Micromagnetic simulations of the vortex core switching threshold vs. spin wave excitation frequency (CCW sense of rotation) for different amplitudes of a simultaneous linear (sub-GHz) gyromode excitation. For CCW excitation multiple switching occurs (exemplified for $A_{\text{gyro}} = 0.38$~mT by the blue dashed line) for all gyromode amplitudes (see text). Furthermore the red cross marks the amplitude and frequency of the simulation snapshot of a ($n$=1, $m$=+2) spin wave mode shown in Fig. 4(a).
    }
    \label{Fig3}
\end{figure}

Additional simulations have been performed with initial vortex core polarization 'up' but with exciting ($n$=1, $m$=+1) spin wave modes by CCW rotating (multi-GHz) field bursts as displayed in Fig.~\ref{Fig3}. The switching threshold without additional gyromode (black squares) clearly shows a minimum at 6.2~GHz which can be explained with the CCW (n=1, m=+1) spin wave eigenmode as shown in \cite{KammererNature}.  The green (yellow, blue) symbols show the spin wave mediated switching threshold for additionally applied 0.25~mT (0.31~mT, 0.38~mT) linear gyromode excitation. Here two additional local minima (at 3.0~GHz and 3.7~GHz) in the switching threshold appear beside the broadband decrease which are explained in the next section. The corresponding experimental diagram cannot be measured because of the onset of multiple switching. This can be understood when comparing the results obtained for the threshold values in Fig.~\ref{Fig2}(b) (CW) and Fig.~\ref{Fig3} (CCW). Both simulations are performed for initial vortex core 'up'. Since the switching threshold for vortex core switching from 'up' to 'down' by CW excitation is significantly lower than the switching threshold for CCW excitation (for corresponding linear gyromode amplitudes as indicated by the dashed blue line in Fig.~\ref{Fig3} exemplified for $A_{\text{gyro}}=0.38~mT$), the vortex core is most likely switched back to its original orientation within the same CCW rotating magnetic field burst. Despite of our time resolution of about 40~ps it is not possible to detect such double switching events during the same ongoing GHz excitation as the strong out-of plane magnetization of the excited spin waves, extended over an area of several 100~nm, hides the tiny point-like out-of-plane magnetization of the vortex core. However, different to the experiment, it is easy to detect multiple switching in micromagnetic simulations by analysing the CCW simulation data. Indeed the multiple switching events are clearly reproduced in our simulations. Although, due to the above mentioned facts, it is not possible to observe the CCW switching behaviour of Fig.~\ref{Fig3} experimentally we were able to assign the second minimum at 3.7~GHz to the appearance of the ($n$=1, $m$=+2) spin wave mode. The characteristic image of this mode was taken experimentally as well as by micromagnetic simulations (c.f. Fig.~\ref{Fig4}(a)) by applying CCW rotating GHz excitation fields but with amplitudes below the CCW switching threshold in order to ensure not to start the multiple switching events.

\section{EXPLANATIONS}
\begin{figure}
    \centerline{\includegraphics[clip,width=0.45\textwidth]{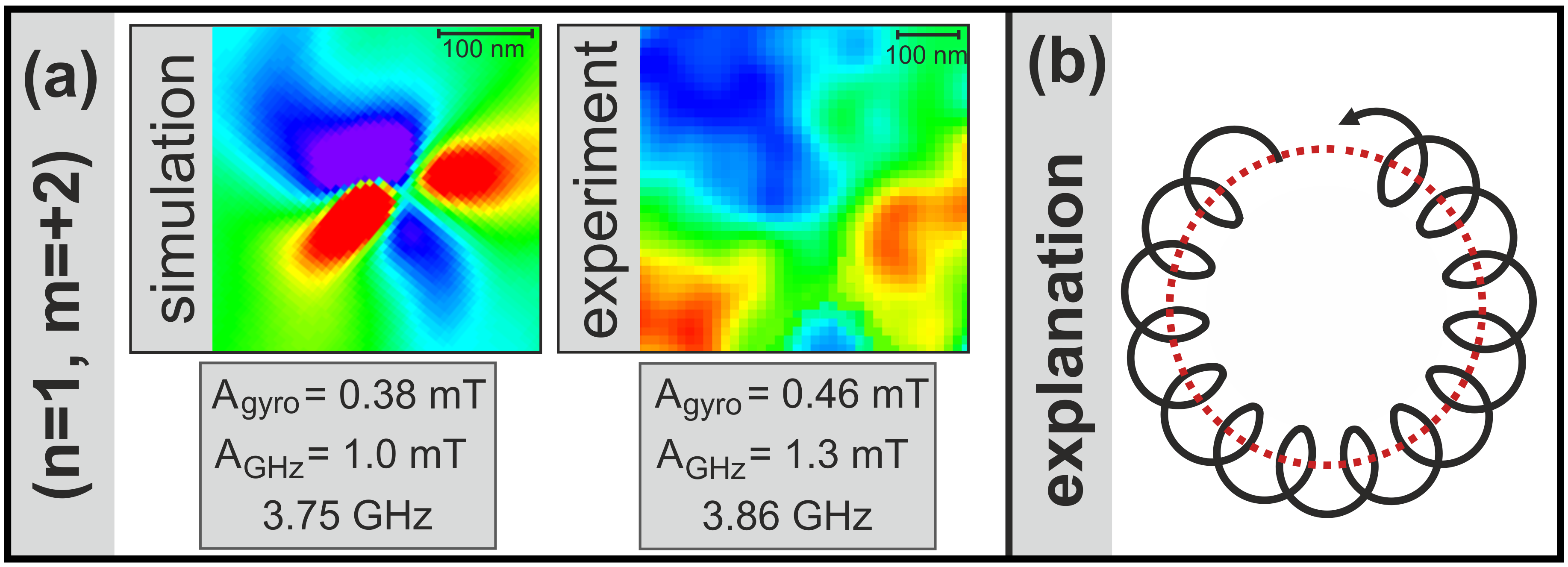}}
    \caption{(a) Comparison of micromagnetic simulations and pump and probe X-ray microscopy experiments to illustrate the appearance of a ($n$=1, $m$=+2) spin wave mode. (b) Sketched vortex core trajectory during simultaneous excitation of the (sub-GHz) gyromode and (multi-GHz) spin wave frequencies.
    }
    \label{Fig4}
\end{figure}

In the following section we explain the results shown above, starting with the broadband decrease of the spin wave mediated switching threshold caused by the simultaneously excited (sub-GHz) gyromode. The vortex core reversal is based on the formation of a vortex-antivortex (VA) pair as demonstrated for both, gyromode \cite{BartelNature,HertelPRL} and spin wave \cite{KammererNature} excitation. Prior to the VA pair formation a 'dip' \cite{NovosadPRB,ArneNature}, i.e., a region with a magnetization direction opposite to the one of the initial vortex core is created by the gyrofield \cite{GuslienkoPRL} of the moving vortex structure. When applying a combined gyromode-spin wave excitation the expected trajectory of the vortex core during this simultaneous excitation is sketched in Fig.~\ref{Fig4}(b) confirmed by our micromagnetic simulations. Note that the radius of the gyromode rotation is about an order of magnitude larger than that of the faster rotation caused by the spin waves \cite{KammererNature}. However, it is obvious that both contributions, (i) the slower gyromode rotation and (ii) the much faster rotation driven by spin wave excitation support the dip formation and in doing so the creation of the VA pair which leads to the vortex core reversal. In that way the switching threshold is lowered significantly as observed experimentally and by micromagnetic simulations. 

In addition to this broadband reduction, additional minima in the spin wave mediated switching threshold were observed experimentally (CW excitation) as well as in our micromagnetic simulations (CW and CCW excitation) caused by the interaction between the gyromode and the spin waves. Beside the well-known spin wave eigenmodes at 5~GHz ($n$=1, $m$=-1, CW) (about 4.5~GHz in the experimental data) and at 6~GHz ($n$=1, $m$=+1, CCW) \cite{KammererNature} additional minima arise (i) at 2.5~GHz for CW excitation and (ii) at 3~GHz as well as at 3.7~GHz for CCW excitation. A closer look at our micromagnetic simulations allows to determine the excited spin wave modes and explains these at first sight unexpected minima. The CW excitation at 2.5~GHz stimulates, by frequency doubling, the ($n$=1, $m$=-1) CW spin wave eigenmode at 5~GHz, but only when the gyromode is excited simultaneously. In the same way the CCW excitation at 3~GHz in combination with gyromode excitation excites the ($n$=1, $m$=+1) CCW spin wave eigenmode at 6~GHz. This frequency doubling effect is a consequence of non-linear interference of both simultaneous excitations and explains the additional minima at 2.5~GHz (CW) and 3~GHz (CCW). However, this cannot explain the minimum in CCW switching threshold at 3.7~GHz.
Here, experiments as well as micromagnetic simulations revealed a different mechanism. A snapshot of the 3.7~GHz CCW mode excited below the switching threshold was recorded by time-resolved scanning transmission X-ray microscopy and is shown in Fig.~\ref{Fig4}(a) together with the result of a micromagnetic simulation. In both cases the azimuthal ($n$=1, $m$=+2, CCW) spin wave eigenmode can be clearly identified by the two minima and two maxima of the out-of-plane magnetization. For symmetry reasons the excitation of ($m$=$\pm$2) spin wave eigenmodes is not possible for a vortex core located at the center of the disc. However, simultaneous excitation of the sub-GHz gyromode shifts the vortex core away from this equilibrium center position leading to the required symmetry breaking and finally to the reversal of the polarization of the vortex core even with ($m$=$\pm$2) spin wave eigenmodes. 

\section{CONCLUSIONS}
Concluding, we have presented changes in the switching threshold which appear when combining the actively excited sub-GHz gyromode with the multi-GHz azimuthal spin waves. These phenomena can be explained by superposition (broadband switching threshold reduction) and by non-linear interference effects (additional minima in switching threshold). These results not only pave the way for more detailed studies of non-linear interactions in vortex structures but might also have a technological impact. Concerning the decrease of the spin wave mediated switching threshold by nearly one order of magnitude is a considerable advantage regarding the realization of a fast VC-MRAM. For the desired smaller element sizes ($< 500$~nm in diameter) and faster switching e.g. by rotating pulse excitation the required field amplitude increases significantly approaching the limit of its technical feasibility. Therefore the in this paper proved immense reduction of the switching threshold amplitudes might be a significant step towards the technological realization of fast VC-MRAMs with current densities in the striplines well below the limit of electromigration and realizable short-pulse power strengths.

\begin{acknowledgments}
Continuous fruitful discussions with Manfred F\"ahnle, MPI Stuttgart, are greatfully acknowledged. We also thank all people involved in the operation of the MAXYMUS scanning X-ray microscope at HZB, BESSY II in Berlin, in particular, Michael Bechtel and Eberhard G\"oring.
\end{acknowledgments}

\end{document}